\documentclass[a4paper,fleqn,usenatbib]{mnras}

\usepackage{mathptmx}
\usepackage[T1]{fontenc}
\usepackage{ae,aecompl}

\usepackage{graphicx}	% Including figure files
\usepackage{amsmath}	% Advanced maths commands
\usepackage{amssymb}	% Extra maths symbols
\usepackage{upgreek}
\usepackage{xcolor}
\title[]{The Non-Uniform Expansion of the Crab Nebula}

\author[T. Martin et al.]{
T. Martin$^{1,2}$\thanks{E-mail: tmartin@cegepgarneau.ca},
D. Milisavljevic$^{3,4}$,
T. Temim$^{5}$,
S. Mandal$^{6}$,
P. Duffell$^{3}$,
L. Drissen$^{1}$,
Z. Ding$^{3}$
\\
$^{1}$D\'epartement de physique, de g\'enie physique et d'optique,
Universit{\'e} Laval, 1045 avenue de la m\'edecine, Qu{\'e}bec,
QC G1V 0A6, Canada\\
$^{2}$C\'egep Garneau, 1660 Boulevard de l'Entente, Qu\'ebec, QC G1S 4S3, Canada\\
$^{3}$Department of Physics and Astronomy, Purdue University, 525 Northwestern Avenue, West Lafayette, IN, USA\\
$^{4}$Integrative Data Science Initiative, Purdue University, West Lafayette, IN 47907, USA\\
$^{5}$Princeton University, 4 Ivy Ln, Princeton, NJ 08544, USA\\
$^{6}$Department of Astronomy, University of Virginia, 530 McCormick Road, Charlottesville, VA, USA}

\date{Accepted XXX. Received YYY; in original form ZZZ}

\pubyear{2023}

\DeclareRobustCommand{\NII}{\textup{N\,\textsc{\lowercase{II}}}}

\DeclareRobustCommand{\Halpha}{\textup{H$\alpha$}}

\newcommand{\vvect}[2]{\begin{pmatrix}#1\\#2\end{pmatrix}}

\hypersetup{draft}
\begin{document}
\label{firstpage}
\pagerange{\pageref{firstpage}--\pageref{lastpage}}
\maketitle

\begin{abstract}
We present extensive proper motion measurements of the Crab Nebula made from Canada-France-Hawaii Telescope MegaPrime/MegaCam images taken in 2007, 2016, and 2019. A total of 19974 proper motion vectors with uncertainty $<10$\,mas\,yr$^{-1}$ located over the majority of the Crab Nebula are used to map the supernova remnant's two-dimensional expansion properties that reflect the dynamics of the original explosion, acceleration of ejecta imparted by spin-down energy from the pulsar, and interaction between the ejecta and surrounding cicumstellar material (CSM). The average convergence date we derive is 1105.5 $\pm$ 0.5 CE, which is 15-35 yr earlier compared to most previous estimates. We find that it varies as a function of position angle around the nebula, with the earliest date and smallest proper motions measured along the equator defined by the east and west bays. The lower acceleration of material along the equatorial plane may be indicative of the supernova's interaction with a disk-like CSM geometry.
Comparing our measurements to previous analytical solutions of the Crab's expansion and our own numerical simulation using the moving mesh hydrodynamics code \texttt{Sprout}, we conclude that the ejecta have relaxed closer to homologous expansion than expected for the commonly adopted pulsar spindown age of $\tau \sim 700$ yr and a pulsar wind nebula (PWN) still evolving inside the flat part of the ejecta density profile. These findings provide further evidence that the PWN has broken out of the inner flat part of the supernova ejecta density profile and has experienced ``blowout''.

\end{abstract}


%%%%%%%%%%%%%%%%%%%%%%%%%%%%%%%%%%%%%%%%%%%%%%%%%%

%%%%%%%%%%%%%%%%% BODY OF PAPER %%%%%%%%%%%%%%%%%%

\section{Introduction}

The first measurement of the expansion of the Crab Nebula was made by
\citet{Duncan1921}, following the first detection of changes in the structure of the supernova remnant made by
\citet{Lampland1921} who used a series of photographs taken with the
40-inch Lowell reflector between 1913 and 1921. The displacement of 12 bright features
among the brightest filaments of the nebula were measured on two
photographs obtained 11.5 years apart at the Mount Wilson reflector
(the first one by Ritchey in 1909 and the other one by Duncan in
1921). A maximum displacement of 0.17\arcsec{}/yr was measured, but the
directions of the proper motion vectors appeared scattered, making it
impossible to determine a unique centre of expansion. A second attempt
made 18 years later by \citet{Duncan1939} based on the measurement of
the displacement of 20 points between the 1909 photograph and a new
photograph obtained in 1939 at the same observatory, clearly showed a
general outward expansion and allowed for the first computation of an
outburst date of 1172 CE, as well as a first estimate of the distance
that yielded 4200 light-years.

Even though the spatial coincidence of the Crab Nebula with a Chinese
guest star observed in 1054\,CE and reported in the translation of the
book 294 of the great Ma-touan-lin collection by
\citet{Biot1846} was first suggested by \citet{Lundmark1921} and
\citet{Hubble1928}, the clear identification of the 1054\,CE event
with the Crab Nebula and a proof that ``a supernova outburst is
accompanied by the throwing off of gaseous shell'' came with the
articles of \citet{Oort1940} and \citet{Duyvendak1942} since the time
coincidence was supported by the now seminal work of
\citet{Duncan1939}. Since the book of \citet{Stephenson2002}, this
explosion date is widely accepted (see
e.g. \citealt{Gaensler2006,Hester2008}).

The remarkable work of \citet{Trimble1968} (thereafter T68) is the
next major step in the study of the expansion of the Crab Nebula. The
use of a filter centered on the \Halpha{} and [\NII{}] $\lambda\lambda$6548, 6583 lines enhanced the visibility of the filaments, permitting the accurate measurement of the proper motion of 132 features, mostly selected in the outer part of the filamentary shell of the supernova remnant (SNR). The proper
motion analysis was based on the comparison of two pairs of plates
taken 14 years apart (1939--1953 and 1950--1964). Along with these
data, the first 3D velocity dataset resulting from the combination of
proper motions and radial velocities (measured via spectroscopy) of 127
features was analysed. The distance estimate of around 2\,kpc was the best to date, and demonstrated that the filamentary
complex was not ``confined to a thin ellipsoid''. Although an outburst date of about 1140 CE could be
calculated since the proper motion vectors do converge generally to
the same centre of origin, the precision of the measurements also
revealed ``some sort of nonuniformity'' and that ``it is, therefore,
only approximately true that the velocity of each filament is
proportional to its distance from the expansion center''.

\citet{Nugent1998} (thereafter N98) used 4 scans of high-resolution
photographs published between 1939-1992 (3 of them found in issues of
Sky \& Telescope and one provided by the Lick Observatory public
information office). The proper motions of 50 features were analysed
to find an outburst date of 1130 CE. From a reanalysis of the same
dataset in a Bayesian framework, \citet{Bietenholz2015} calculated a
value of 1091\,CE, somewhat nearer the supernova explosion date of
1054\,CE.

\citet{Wyckoff1977} (thereafter WM77), N98,
\citet{Kaplan2008} (thereafter K08) and \citet{Bietenholz2015}  (thereafter BN15), all draw similar conclusions,
although the location of the explosion centre 
differs by a few arcseconds (N98, K08). From the analysis made by
T68 and N98 (reanalysed later by
\citealt{Bietenholz2015}), the homologous expansion of the outburst appears to be well established. However, in all these works, only a few filaments,
mostly located on the border of the remnant (where the proper motion
is larger), were considered. The one exception is work by \citet{Rudie2008}, who measured proper motions for 35 locations in the northern filamentary jet. In that case, an age of $1055 \pm  24$ CE was estimated, which they interpret as evidence that the jet experienced less outward acceleration from the central pulsar's rapidly expanding synchrotron nebula when compared to the main body of the remnant.

In this paper we conduct our own proper motion measurements of the Crab Nebula. Our original motivation was to investigate in detail the validity of the
homologous expansion model (see for example
\citealt{Wyckoff1977,Nugent1998,Bietenholz2015}) described below
(eq.~\ref{eq:expansion_model}) and used by \citet{Martin2021} to
project the $\sim$400,000 radial velocity points obtained with SITELLE \citep{Drissen19} in the Euclidean space. This model is simply based on an unaccelerated
motion of the ejected material which means that the velocity of the
material, i.e. its proper motion $\mu$ in the celestial plane, is
strictly proportional to its distance to the centre of expansion $r$,
\begin{equation}
  \label{eq:expansion_model}
  \mu = \frac{r}{t}\;,
\end{equation}
where $t$ is a measure of the expansion age (WM77). If
the material is truly not accelerated this expansion age should
correspond to the time elapsed since 1054\,CE. The idea behind
\citet{Martin2021} reconstruction is that, from this model, and
measuring the mean expansion factor at a certain time, one can infer
the velocity in the celestial plane from the distance of the material
to the expansion centre. From there, each radial velocity point can be
combined to an estimated velocity in the celestial plane to obtain a
set of 3D velocity vectors that can then be used to estimate their
position in space knowing the outburst date and the distance to the
Crab. However, if this expansion is not homologous, i.e. if the
material is accelerated in some directions, the overall shape of
the SNR computed by \citet{Martin2021} may be substantially
different.

Obtaining a precise and complete mapping of the expansion may furthermore offer the possibility to study the acceleration inhomogeneities of the supernova material which may give insights on the dynamics in play since the initial outburst.

\section{Data}

\begin{figure*}
  \includegraphics[width=\linewidth]{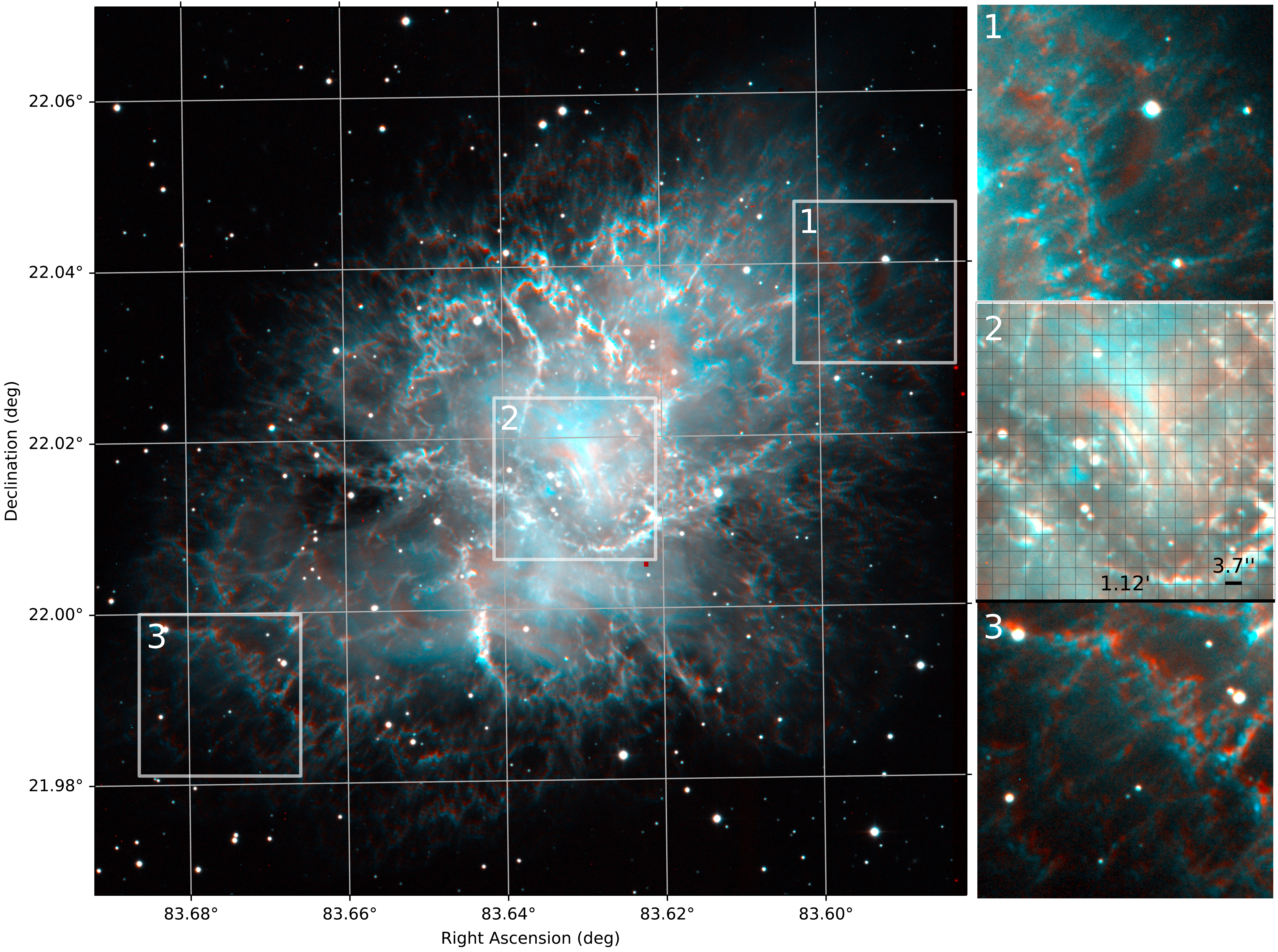}
  \caption{Combination of the images taken in 2007 (red) and 2016
    (blue) with 3 zoomed-in regions on the right. The expansion of the
    supernova remnant is conspicuous. We added a grid on region 2 to
    show the size of the tiles compared to compute the proper motion
    (see section~\ref{sec:propermotion}).}
  \label{fig:contrast-gimp}
\end{figure*}

We have selected three sets of observations taken with Canada-France-Hawaii Telescope (CFHT) MegaPrime/MegaCam instrument through two similar red filters (r and rS) in 2007, 2016 and 2019 (see Figure~\ref{fig:contrast-gimp}). Even if multiple exposures were taken we have decided not to combine them so as to avoid additional distortions. In the 2019 dataset we have
selected the image number 2434157 because the whole remnant fits in
one chip. In the 2007 set, the Crab Nebula is always split between two
different chips (see Figure~\ref{fig:chips}). We therefore selected
two different images (906610 and 905816) that show a small
overlap. The image quality of the two sets of images is comparable and
found to be around 0.8\arcsec{} (see Table~\ref{tab:data}).

\begin{table}
  \centering
  \caption{Images compared to compute the proper motion. They were all
    obtained with MegaPrime/MegaCam at the Canada-France-Hawaii
    Telescope (CFHT).}
  \begin{tabular}{l|cccc}
    \label{tab:data}
    Image number&Chip&Exposure date&Filter&FWHM (``) \\
    \hline
    2434157&22&July 9, 2019&r&0.7\\
    1892362&22&February 7, 2016&r&1.0\\
    906610&20&April 13, 2007&rS&0.9\\
    905816&21&April 13, 2007&rS&0.7\\
    \hline
    \multicolumn{5}{l}{Plate scale: 0.1869\,\arcsec{}/pixel}\\
    \multicolumn{5}{l}{Time span of the 2016--2007 set: 8.83 yr (3224 jd)}\\
    \multicolumn{5}{l}{Time span of the 2019--2007 set: 12.24 yr (4472 jd)}\\
    \hline
  \end{tabular}
\end{table}

\section{Proper motion computation method}
\label{sec:propermotion}
The general idea is to estimate the displacement (in pixels) of the
features between 2007 and 2016 or 2019 in small 20 pixel-wide boxes
(3.7\arcsec{}$\times$3.7\arcsec{}) which can then be translated to a
proper motion. Therefore, the compared region must be perfectly
aligned and, if possible, free of any significant distortion. The
images are thus first registered and aligned (see
section~\ref{sec:align}). The displacement vectors are computed at
different positions on the Crab Nebula (see
section~\ref{sec:pmcompute}) and corrected for the point spread
function (PSF) asymmetry and the mean proper motion in this field (see
section~\ref{sec:pmcorrect}). Finally the proper motion in celestial
coordinates can be calculated (see section~\ref{sec:pmproject}).

\subsection{Step 1: registration and alignment of the images}
\label{sec:align}

Each image was first registered independently using the Gaia DR3
\citep{GaiaCollaboration2016,Collaboration2022} catalog (see
Figure~\ref{fig:chips}). 264 stars were selected in the vicinity of
the Crab Nebula. Stars with a large proper motion were removed from
the list of stars used. In order to mitigate the effects of large
scale distortion the registration was made with a Simple Imaging
Polynomial \citep{Shupe2005} of degree 3 for the images taken in 2007
and of degree 2 for the images taken in 2016 and 2019 which tends to
display better optical properties. This degree was found to be the largest needed for an adequate Gaussian distribution of the residuals and
small enough to avoid overfitting.

Registration errors are generally below 0.1\arcsec{} ($\approx$0.5
pixel, see Figure~\ref{fig:chips}), which translate to an uncertainty
of 8.17 to 11.3\,mas/yr when comparing images separated by 8.83 and 12.24
years (see Table~\ref{tab:data}).

Then the two 2007 fields are projected on the 2019 field (which is
used as the reference field in the whole analysis since it is the only
one where the entire Crab fits within a single Megacam chip). We used the
Python package \texttt{reproject} \citep{Robitaille2020} for this
purpose. As the registration process does not involve non-linear
geometric correction, the projection is robust and does not induce any
small scale distortion. We used a flux-conserving spherical polygon
intersection algorithm for the reprojection. The projected images were
used to produce the layered image shown in
Figure~\ref{fig:contrast-gimp}.

\subsection{Step 2: proper motion computation}
\label{sec:pmcompute}

Once the compared images are properly aligned we can obtain the proper
motion in right ascension and declination at any given position by
computing the displacement of the image in a small
20$\times$20\,pixel$^2$ tile (3.7\arcsec{}$\times$3.7\arcsec{}) around
the chosen coordinates.

\begin{figure}
  \includegraphics[width=\linewidth]{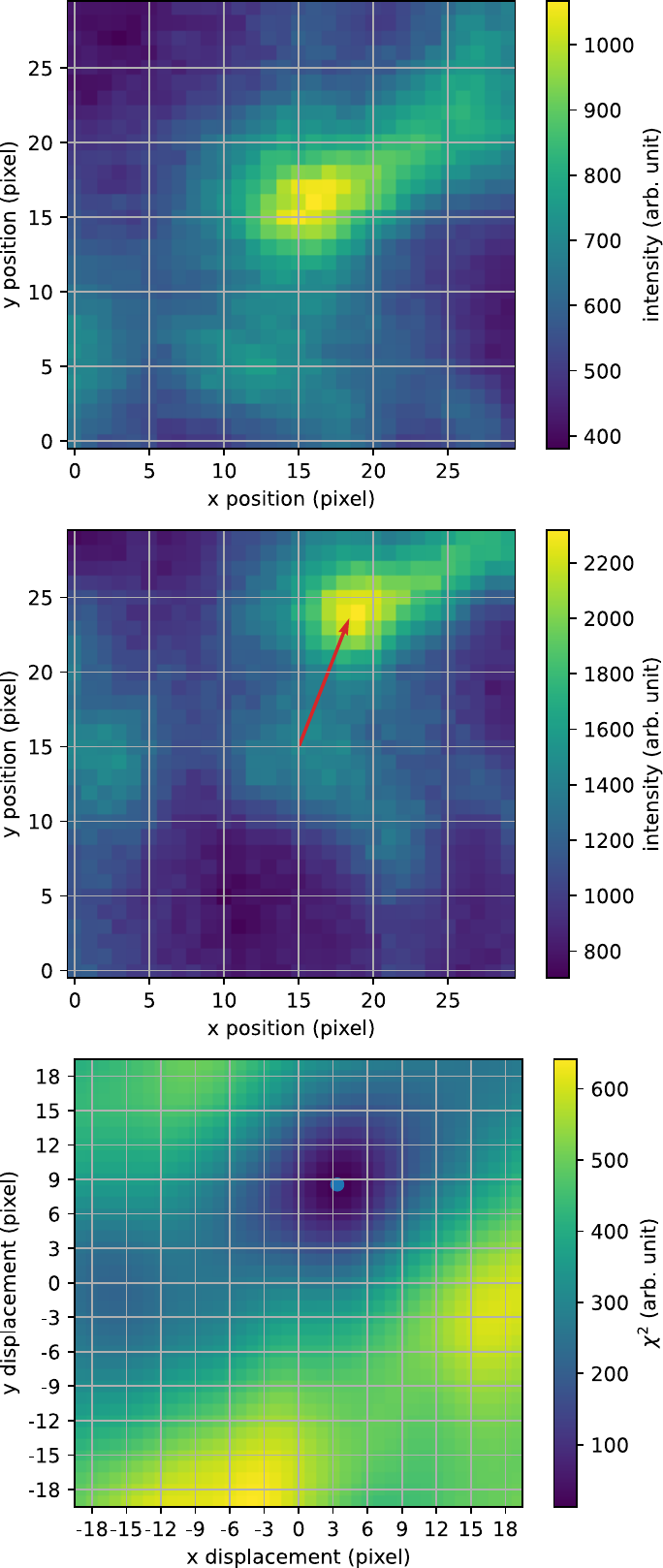}
  \caption{\textit{(Top and centre panels)} Example of two compared
    tiles taken at the same celestial coordinates on the 2007 and 2016
    images. The red arrow indicates the estimated displacement
    vector. \textit{(Bottom panel)} Calculated $\chi^2$ matrix of all
    the possible shifts tried along the x and y axes. The position of
    the estimated best shift is located with a blue dot.}
  \label{fig:algo}
\end{figure}

For this purpose, instead of directly using  a
FFT-based cross-correlation algorithm, which is faster but has numerous border effects on
small regions, we explore a range of shift possibilities
along the x and y axes on one tile and keep the one that shows the
highest correlation with the reference tile. The highest
correlation is found when the distance between both the shifted tile
and the reference tile is minimal in least square terms. The loss
function we try to minimise can be expressed as a
$\chi^2({\Delta x, \Delta y})$ of two parameters: $\Delta x$ and
$\Delta y$, resp. the x and y shift of the tile:
\begin{equation}
  \label{eq:algo}
  \chi^2({\Delta x, \Delta y}) = \sum (I_{\text{shift}}({\Delta x, \Delta y}) - I_{\text{ref}})^2\;,
\end{equation}
where $I_{\text{shift}}({\Delta x, \Delta y})$ is the shifted tile and
$I_{\text{ref}}$ is the reference tile (see
figure~\ref{fig:algo}). Note that, before being compared the tiles are
normalised and clipped at their respective 25th and 75th percentiles
to mitigate the effects due to a different sky transparency or the
variation of the sky background.

In principle a minimisation algorithm could be used directly. But,
because non-integer shifts rely on an interpolation algorithm that
produces small distortions, we want to limit the shifts to integer
values. Also, in order to avoid a local minimum, we use
a brute force minimisation. Shifts from -3.7\,\arcsec{} to
+3.7\,\arcsec{} (approx. -2900 to 2900\,km/s) in steps of 1 pixel
0.037\arcsec{} (or 0.187\, arcsec i.e. approx. 150\,km/s) in both x
and y directions were attempted. This process yields a $\chi^2$ matrix where the global
minimum can be securely found (see figure~\ref{fig:algo}). The
precision of this minimisation can then be greatly enhanced by
interpolating the $\chi^2$ matrix around the minimum with a 3rd order
polynomial, the subpixel precision minimum being the minimum of this
polynomial.

We have tested the precision of this algorithm on simulated data -- a
large gaussian (with a fwhm of 10 pixels) shifted by an arbitrary
value -- and we found that, in the ideal noiseless case, the shift of
the gaussian could be retrieved with a precision of $5\times10^{-4}$
pixel. It is clearly not the precision that can be obtained on real
data obtained under different conditions on a changing object, but the exercise demonstrates convincingly that it won't be limited by the algorithm.

Interestingly enough we found that the algorithm developed for this
study was in fact very similar to the processing involved in Particle
Image Velocimetry (PIV) used in the field of fluid mechanics to map
fluid flows (see e.g. \citealt{Thielicke2014,Adrian2003,Keane1990})
with the difference that the images analysed in PIV are made of
numerous point-like objects. The method used to
determine the most probable shift between two images generally relies on the
maximisation of a cross-correlation function\footnote{In the PIV cases,
  equation~\ref{eq:algo} would be replaced with \begin{equation}
    C({\Delta x, \Delta y}) = \sum I_{\text{shift}}({\Delta x, \Delta
      y}) I_{\text{ref}}\;.
  \end{equation}}, while our method relies on the minimisation of the quadratic
distance. We tried both methods and found that the returned matrices (a
cross-correlation matrix in the PIV case and a $\chi^2$ matrix in our
case) were in fact very similar. The precision of both methods on our
simulated data was nearly the same, though our algorithm
yields a marginally better precision on a noiseless set of data.

The displacement vectors obtained with this algorithm should be, in
principle, further corrected for residual local distortions
unaccounted for during the registration step. But the relatively low
precision on the estimation of the position of the individual stars
compared to the quality of the registration makes this step
unnecessary since the estimation of the local distortion field must
logically depend on a small number of stars. However, the precision of
the estimation based on e.g. 5 stars appears too low when compared
to the precision of the large scale registration. We found that, a
correction based on a small number of stars was in fact increasing the
scattering of the data. We thus decided to rely on a much larger set
of stars.

% ...............

\subsection{Step 3: corrections for systematics}
\label{sec:pmcorrect}

Systematic biases in the calculated proper motions can greatly affect
the position of the derived expansion center (see
section~\ref{sec:whole_nebula}). Over a $\sim 10$ yr span, a
relative shift of 0.1\,pixel (19\,mas) between the compared images
translates to a proper motion shift of 1.9\,mas/yr and thus a final
shift of the calculated expansion center of 1.9\arcsec, given an
expansion age of $\sim 1000$ years.

\begin{figure}
  \includegraphics[width=\linewidth]{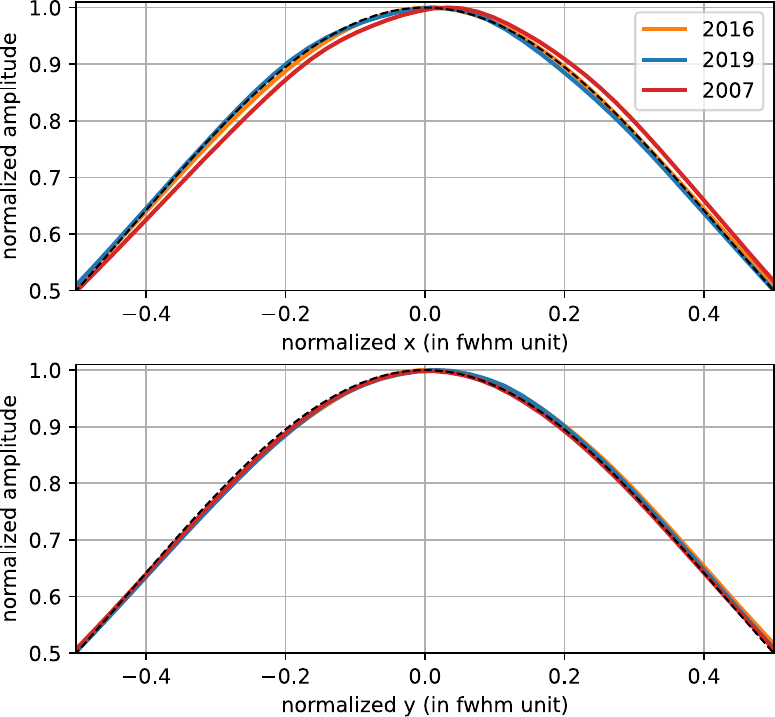}
  \caption{Normalized computed PSF of the stars in the compared
    images. Only the superior part is drawn to better show the
    assymmetry of the 2007 PSF along the x axis which leads to a
    non-negligible systematic on the computed proper motion along the
    RA axis.}
  \label{fig:psf}
\end{figure}

Multiple issues contribute to a systematic shift in the measured
proper motion.
\begin{itemize}
\item Registration biases. In our case, the mean registration error is
  never higher than 0.02 pixel, i.e. 4\,mas (see
  section~\ref{sec:megacam_images}) and was direclty subtracted from
  the measured proper motion. But the uncertainty on this bias is
  comparatively higher and it must be accounted for in the error
  budget.  Considering one standard deviation as the maximum
  systematic registration error, registration biases could contribute
  to a systematic shift of 0.07 pixel ($\approx 12$\,mas) between the
  compared
  images.
\item Mean proper motion of the registration stars. As the stars are
  aligned during the registration, the mean proper motion of the stars
  used for the registration is added to the computed proper
  motion. Using the Gaia catalog, we can estimate it and substract its
  contribution (see Table~\ref{table:shift}).
\item Point Spread Function (PSF) Asymmetries. Differences in the
  optics used, guiding errors, atmospheric conditions and other
  factors can modify the shape of the PSF. Any asymmetry will result
  in a systematic bias of the measured centroid which will eventually
  produce an additional registration shift. We have normalized and
  stacked together the stars used for the registration on a grid
  oversampled by a factor of 10 for every image of our dataset (see
  Figure~\ref{fig:psf}). The resulting high SNR PSFs have subsequently
  been fitted with the same algorithm used during the registration in
  order to estimate the relative error made on the position of the
  centroid between the compared images.
\item Parallax. At a distance of 2\,kpc, the maximum parallax is
  0.5\,mas which can be safely neglected given the magnitude of the
  preceding contributions.
\end{itemize}

\begin{table}
  \centering
  \caption{Systematics estimated (and corrected) caused by the mean
    proper motion of the reference stars and the PSF assymetries.}
  \begin{tabular}{l|cccc}
    \label{table:shift}
    Sets&\multicolumn{2}{c}{Mean proper motion}&\multicolumn{2}{c}{PSF Asymmetry}\\
        &$\Delta\alpha$&$\Delta\delta$&$\Delta\alpha$&$\Delta\delta$\\
        &mas&mas&mas&mas\\
    \hline
    2007--2016&5.15(16)&9.88(10)&13.47(84)&-1.02(85)\\
    2007--2019&7.15(17)&13.71(11)&19.25(73)&-2.52(73)\\
    \hline
  \end{tabular}
\end{table}

\subsection{Step 4: projection to celestial coordinates}
\label{sec:pmproject}

The raw output of our algorithm is an estimate of the pixel
displacement $(\mu_x, \mu_y)$ of a small region centered at the pixel
coordinates $(x, y)$ between two images obtained at two different
times $t_1$ and $t_2$. From there, given the parameters of a World
Coordinate System (WCS) \citep{Greisen2002} that transform the pixel
coordinates to celestial coordinates, the proper motion
$(\mu_\alpha, \mu_\delta)$ is obtained as follow:
\begin{gather}
  \vvect{\alpha}{\delta} = \text{WCS}\vvect{x}{y}\\
  \vvect{\mu_\alpha}{\mu_\delta} = \left(\frac{1}{t_2- t_1}\right)\text{WCS}\vvect{x + \mu_x}{y + \mu_y} - \vvect{\alpha}{\delta}\;,\\
  \mu_\alpha^\star = \mu_\alpha \cos(\delta)
\end{gather}

With two sets of data to compare (the 2019-2007 set and the
2016-2007 set) that should give the same values, we
can estimate the uncertainties on our method. If we look at
the difference between the proper motion vectors obtained from the
2019 set and the 2016 set computed over the whole SNR (see
section~\ref{sec:whole_nebula}), we see that the median difference is
smaller than 2\,mas/yr and the standard deviation is around
13\,mas/yr. Given the fact that the uncertainty found when comparing
our data with that of T68 (see
section~\ref{sec:trimble}) is around 10\,mas/yr (when combining the
proper motion measured with the 2007--2016 and the 2007-2019 sets), we
conclude that the measured uncertainty of 13\,mas/yr is a good
estimate of the uncertainty of the proper motion vectors computed when
comparing two images. When combining both sets, this error falls
to 9.1\,mas/yr. The complete error budget of our data is presented in
Table~\ref{table:budget}.

\begin{table}
  \centering
  \caption{Errors budget on the position and the magnitude of the
    calculated proper motion vectors. The combined error refers to the
    fact that the proper motion further analysed in this study is the
    mean of the proper motion computed for the 2007--2016 and the
    2007--2019 sets.}
  \begin{tabular}{l|cccc}
    \label{table:budget}
    Sets&\multicolumn{2}{c}{Proper motion}&\multicolumn{2}{c}{Position}\\
        &\multicolumn{2}{c}{mas/yr}&\multicolumn{2}{c}{mas}\\
        &Systematic&Random&Systematic&Random\\
    \hline
    2007--2016&1.4&13&6.1&6.1\\
    2007--2019&1.0&13&6.1&6.1\\
    \hline
    Combined&0.9&9.2&4.3&4.3\\
  \end{tabular}
\end{table}

\section{Comparison with Trimble (1968)}
\label{sec:trimble}

\citet{Trimble1968} reported in her paper a table (Table II) of the
132 proper motion vectors she used in her analysis as well as their
positions with respect to the central double star previously used by
\citet{Duncan1921} and later by N98 and K08. This star was well chosen as a reference as it
displays a very small proper motion (see Table~\ref{table:duncan}).
\begin{table}
  \centering
  \caption{\citet{Duncan1921} Reference star coordinates and proper motion from Gaia DR3 \citep{GaiaCollaboration2016,Collaboration2022}}
  \begin{tabular}{l|c}
    \label{table:duncan}
    Designation&Gaia DR3 3403818176867563264\\
    \hline
    $\alpha$ ($^{\circ}$)&83.6341075 $\pm$ 4.15$\times 10^{-5}$\\
    $\delta$ ($^{\circ}$)&22.0155339 $\pm$ 3.45$\times 10^{-5}$\\
    $\mu_\alpha$ (mas/yr)&2.291 $\pm$ 0.202\\
    $\mu_\delta$ (mas/yr)&-5.382 $\pm$ 0.130\\
    \hline
  \end{tabular}
\end{table}

\begin{figure}
  \includegraphics[width=\linewidth]{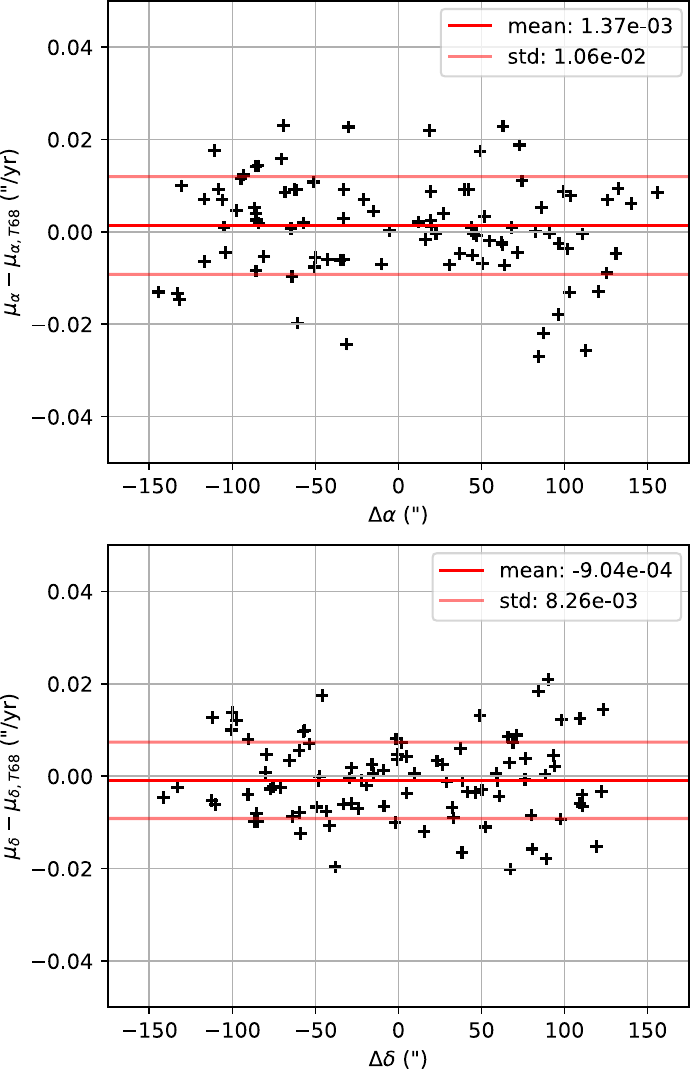}
  \caption{Difference with T68 on the measured proper motion along
    right ascension (top) and declination (bottom). The mean and
    standard deviation are indicated in each panel.}
  \label{fig:trimble_diff}
\end{figure}

We may hence compare the proper motion we compute from our data at the
positions chosen by T68. But we cannot simply look at
the exact same coordinates since the Crab has since expanded and the
regions considered by T68 are now a little farther
away. We therefore estimated the actual coordinates of these regions
from the measured proper motion, computed with respect to the
estimated position of the reference star in 1966. The results of this
comparison are summarised in Figure~\ref{fig:trimble_diff}.
Note that this comparison is only correct from a statistical point of
view because we did not try to find the exact same positions as the
one used by T68. The idea behind this comparison is to
rule out the possibility of any significant systematic error in our
method and to obtain an upper limit on its precision.

The difference between the proper motions measured and those of
T68 show a deviation $\lesssim 1.6$\,mas/yr with an
uncertainty of 10\,mas/yr which is very comparable to the
uncertainties quoted in section~\ref{sec:pmproject}.

\section{Analysis of the proper motion of the whole nebula}
\label{sec:whole_nebula}
A total of 22957 proper motion vectors were computed at positions
defined from a grid covering the whole remnant with a step size of 7
pixels (1.3\arcsec) from which only positions in the filamentary
regions (derived from the \Halpha{} map obtained with SITELLE and
analysed in \citealt{Martin2021}) and sufficiently far from the stars
(Gaia DR3 catalog \citealt{GaiaCollaboration2016,Collaboration2022}) were
considered.

\begin{figure*}
  \includegraphics[width=\linewidth]{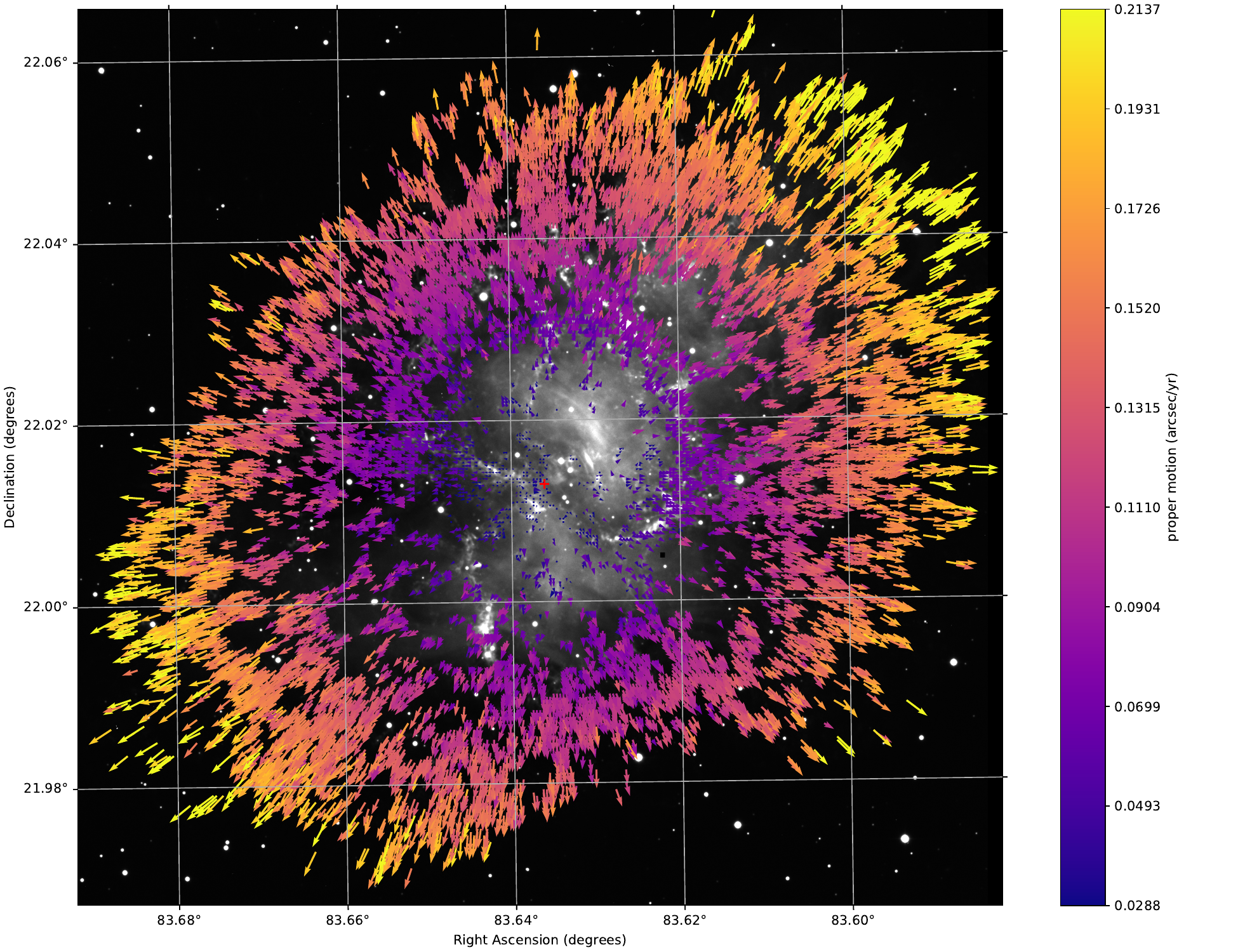}
  \caption{Proper motion vectors measured on the filamentary regions
    of the Crab Nebula from our
    Megacam images. For a more readable representation, only half
    of all 19974 vectors are shown. The background
    image is the 2019 observation. The arrows point to the
    position the material will occupy in 50 years from now. 
    The red cross is our derived centre of expansion.}
  \label{fig:pm_whole_field}
\end{figure*}

We applied two filters to remove vectors associated with weak emission along the outer parts of the remnant and image defects (e.g. hot pixels). The first one is based
on the comparison of the proper motion obtained from the 2016-2007 and
2019-2007 sets. The 2359 vectors deviating by more than 5 times the
uncertainty calculated in section~\ref{sec:pmproject} were
removed. The second filter is based on the hypothesis that the
computed vectors should roughly follow a homologous law
($\mu \propto r$). 624 vectors deviating by more than 5 times the
standard deviation of the distribution around a fit to this law were
also removed. Following this, 19974 proper motion vectors were
retained. The proper motion map obtained is shown in
Figure~\ref{fig:pm_whole_field}.

\subsection{Expansion center and expansion age}
\label{sec:expansion_center}

Two methods have been used by different authors to calculate the
expansion age of the SNR and the position of its expansion
centre. Both methods are based on a constant velocity uniform expansion model (see
equation~\ref{eq:expansion_model}), which has been largely proven to
be inappropriate since the resulting outburst date is always calculated to be
later than the canonical 1054\,CE by a hundred years (see
\citealt{Duncan1921}, \citealt{Bietenholz2015} and the more recent
calculated outburst dates reported in Table~\ref{tab:centers}).

The first method, used by T68 consists in computing
the time of closest approach in the least-square sense. Given the
position and proper motion of some vectors, one can calculate backward
their position and find the moment where they cover the smallest
region in the sky. The mean position of the vectors at this time gives
the position of the expansion center. We have written this algorithm
and compared its results to the results obtained on the same data by
T68 to verify that our implementation was
correct. 

The second method, used by WM77 and N98, relies on the
least-square fit of an unaccelerated expansion model (see
equation~\ref{eq:expansion_model}) relating the proper motion
($\mu_\alpha$, $\mu_\delta$) to the right ascension and declination
coordinates with respect to the Duncan reference star ($\Delta\alpha$,
$\Delta\delta$)
\begin{gather}
  \label{eq:lsqfit}
  \mu_{\alpha} = t^{-1} \Delta\alpha\;,\\
  \label{eq:lsqfit2}
  \mu_{\delta} = t^{-1} \Delta\delta\;.
\end{gather}

\begin{figure}
  \includegraphics[width=\linewidth]{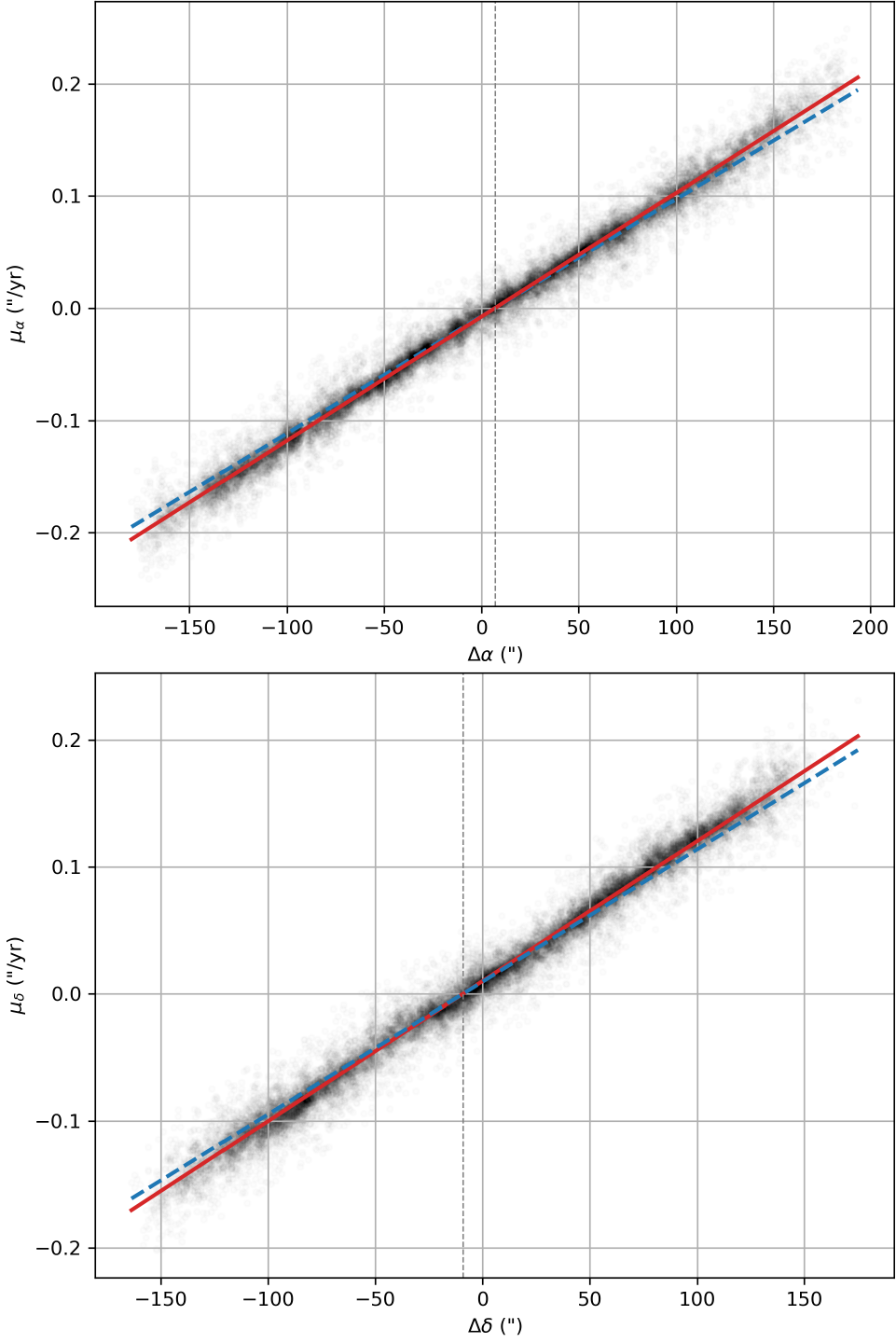}
  \caption{Proper motion vs position with respect to the
    \citet{Duncan1921} reference star along right ascension (top) and
    declination (bottom). The red line shows the result of a least
    square fit using the unaccelerated model of
    equations~\ref{eq:lsqfit} and~\ref{eq:lsqfit2}. 
    The blue line illustrates the result of an unaccelerated expansion
    since 1054\,CE.}
  \label{fig:pmradec}
\end{figure}

The resulting fit on our data is shown on Figure~\ref{fig:pmradec}.

\begin{figure}
  \includegraphics[width=\linewidth]{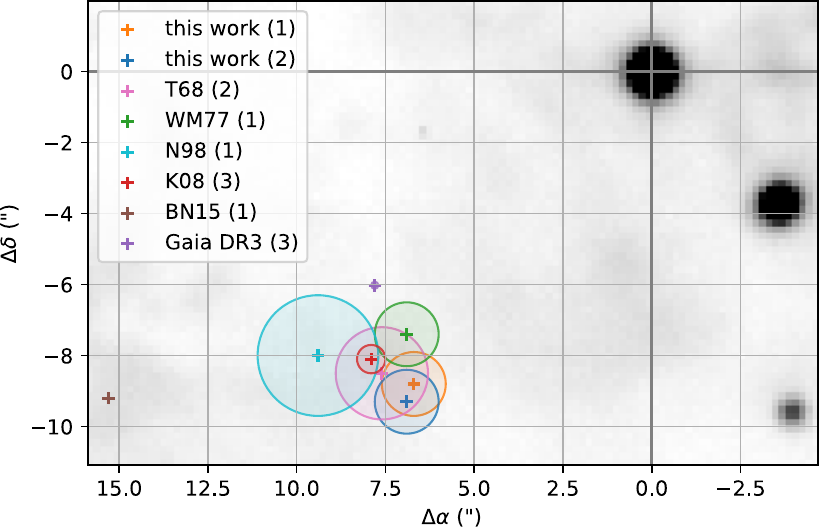}
  \caption{Map of the expansion centres computed by different authors
    and reported in Table~\ref{tab:centers} including the position
    computed in this work with (1) a fit on a proper motion vs
    position graph (chosen as a reference in this article) and (2) the closest approach method. 1-$\upsigma$ uncertainties are shown as circles. The arrow
    and the dashed gray line indicate the direction of the proper
    motion of the Crab pulsar computed by K08. The positions are shown with respect to the \citet{Duncan1921} reference star located at (0,0).  
    }
  \label{fig:map}
\end{figure}

The results obtained with both methods are presented in
Table~\ref{tab:centers} and Figure~\ref{fig:map}. The uncertainties on
these values were computed with a bootstrap method in which the
parameters are recomputed a thousand times with the data points each
time shifted by a random value sampled from a Normal distribution
reflecting the systematics and random uncertainties quoted in
Table~\ref{table:budget}.

We found nearly the same positions of the expansion center with both
methods (see Table~\ref{tab:centers}), but our calculated outburst
date is very different from the date obtained with the other
method. Note that the same discrepency was already observed by WM77
when they applied the second method to the data obtained by T68. We
have decided to retain the results obtained with the fit method since
it is much easier to implement and seems, in this way, more
statistically robust. As an example, with the closest approach method,
the region occupied at the closest approach time is, in fact, very
large (the 1-$\upsigma$ diameter of this region is around 20\,\arcsec)
because the proper motion uncertainty projected a thousand years back
is large. This added step of backward projection may influence the
statistics of the computed expansion date in some ways that are more
difficult to control. Given the fact that, in principle, both methods
should give the same results, we favor the simplest of both.

\begin{table}
  \centering
        \caption{Expansion centres computed by different authors (and
    compiled by N98 and K08) including the positions computed in this
    work with (1) a fit on the proper motion vs position graph
    (chosen as a reference in this article) and (2) the closest
    approach method. The positions are calculated with respect to the
    \citet{Duncan1921} reference star. See Figure~\ref{fig:map} for a
    map of these positions. The resulting outburst date considering an
    unaccelerated expansion is also reported.}
  \begin{tabular}{l|ccl}
    \label{tab:centers}
    Reference&$\Delta\alpha$&$\Delta\delta$&Date\\
    method& (\arcsec)& (\arcsec)\\
    \hline
    \textbf{This work}$^1$&\textbf{6.9(0.8)}&\textbf{-9.3(0.9)}&\textbf{1105.5(0.5)}\\
    This work$^2$&6.7(0.8)&-8.8(0.9)&1140.5(1.4)\\
    T68$^2$&7.6(1.3)&-8.5(1.1)&1140(15)\\
    WM77$^1$ (T68 data)&6.9(0.9)&-7.4(0.9)&1120(7)\\
    N98$^1$&9.4(1.7)&-8.0(1.3)&1130(16)\\
    BN15$^1$ (N98 data)&15.3&-9.2&1091(34)\\
    BN15 (synchrotron)&--&--&1255(27)\\
    \hline
    K08$^3$&7.9(0.4)&-8.1(0.4)&\\
    Gaia DR3$^3$&7.8(0.1)&-6.02(0.06)&\\
    \hline
    \hline
    \multicolumn{4}{l}{$^1$ Fit on a proper motion vs position graph}\\ \multicolumn{4}{l}{(retained as a reference).}\\
    \multicolumn{4}{l}{$^2$ Closest approach method.}\\
    \multicolumn{4}{l}{$^3$ Computed position of the pulsar in 1054\,CE based on}\\
    \multicolumn{4}{l}{its measured proper motion.}
  \end{tabular}
\label{tab:centers}
\end{table}

\subsection{2D expansion properties}

\begin{figure*}
\centering
  \includegraphics[width=0.75\linewidth]{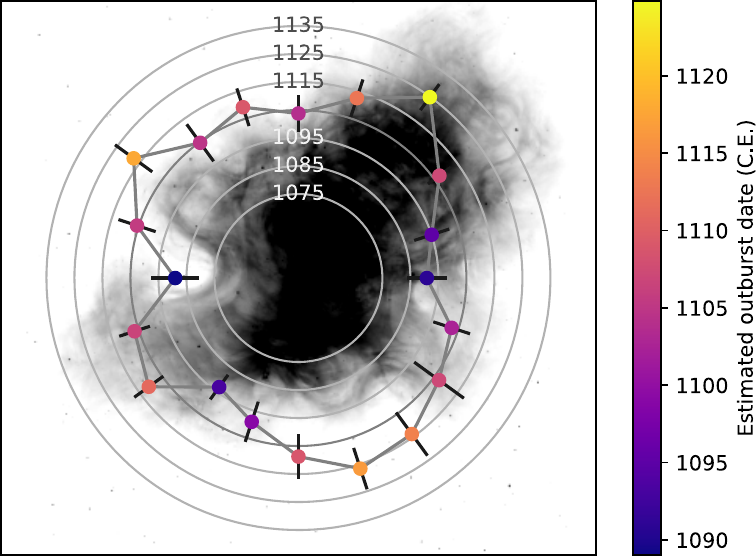}
  \caption{Inferred outburst date as a function of position angle. Each point is the outburst date computed by taking the points contained in small angular wedges around the nebula. The uncertainty is indicated by black error bars. The darker grey line demarcates the radial distance corresponding to our inferred outburst date using all data. The background image is a James Webb Space Telescope continuum map published in \citet{Temim24}.}
  \label{fig:pa-pa}
\end{figure*}

The quality of our measurements allowed us to go beyond a straightforward average of the proper motions of all data and investigate how expansion properties of the Crab may change as a function of position angle from the centre of expansion. We divided the remnant into 20 angular wedges originating from the centre of expansion, then re-ran the same procedure outlined in section \ref{sec:whole_nebula} using the data from these wedges. The results are shown in Figure~\ref{fig:pa-pa}. There is considerable difference in the inferred outburst date among the radial segments. The youngest ages are found in the north and south, whereas the oldest ages consistent with the least amount of acceleration are found in the east and west. The maximum age of 1115 C.E. is in the northwest region  where ``breakout'' of the radio synchrotron nebula is observed \citep{SH1997}. Clearly, this morphology was imprinted on the nebula's kinematics by the strong relativistic wind along the pulsar's rotation axis. 

The smallest proper motions are measured along the equator defined by the east and west bays (around position angles 90 and 270 degrees). These small proper motions are consistent with radial velocity measurements from published optical spectra showing conspicuously reduced ejecta velocities ``pinched'' in these regions \citep{Fesen1997}. This arrested expansion, along with enhanced helium abundances in filaments in these regions \citep{MacAlpine1989,Fesen1997,Satterfield12}, may be indicative of interaction with a concentration of circumstellar material (CSM) that impeded the expansion \citep{Fesen1992,Smith13}. 

The presence of a pre-SN CSM that the blast wave and ejecta ran into and interacted with shortly after core collapse has been proposed to help explain the unusually high luminosity of the original supernova (-18 mag; \citealt{Stephenson2002}) that produced the Crab Nebula reported in historical records, which stands in contradiction with the surprisingly low kinetic energy ($\approx 7 \times 10^{49}$\,erg) inferred from the total ejecta observed now (see, e.g., \citealt{CU00}). Nonetheless, other interpretations for the high luminosity exist (see, e.g., \citealt{Tominaga13, Omand2024}), as well as an interpretation that the east-west filaments appear particularly prominent due to projection effects \citep{Hester1995}. While other bay-like structures are observed around the perimeter of the PWN that may suggest that confinement is mainly attributable to prominent ejecta filaments \citep{Temim24}, Figure~\ref{fig:pa-pa} shows that the acceleration of material is lowest along the east and west bays. Some 3D magnetohydrodynamic models predict an elongation of the PWN along the pulsar’s rotation axis \citep{Olmi2016}, that may result in anisotropic acceleration of the ejecta filaments. However, we measure less accelerated material along the east-west direction and not along the torus axis, as would be expected in that scenario. The observed “pinch” along the equatorial region with the more accelerated north and south lobes seen in Figure~\ref{fig:pa-pa} could instead be explained by either an asymmetric explosion or more likely a pre-existing disk-like CSM morphology. A definitive proof for significant CSM material is still lacking, but future analysis of the kinematic and chemical properties of the Crab Nebula in 3D enabled by multiple epochs of SITELLE observations could provide the necessary confirmation.

\section{Numerical modeling}

From the assumption that the supernova outburst took place in 1054 C.E., the actual proper motion is too fast to be explained with a ballistic, homologous expansion ($v=r/t$, see Figure 6). Instead, the younger inferred explosion date may be explained by taking into account the wind driven by the central pulsar, which accelerates the ejecta. Assuming uniform density in the interior of the stellar ejecta, the wind-accelerated ejecta can be shown to evolve in size as $r \propto t^{6/5}$ (using dimensional analysis; also see \citealt{Chevalier77}) instead of $r \propto t$ (as for free expansion). 
This will lead to an expansion velocity $v$ greater than $r/t$:
\begin{equation}
    v = \dot{r} \propto \frac{\text{d}(t^{6/5})}{\text{dt}} = \frac{6}{5} t^{1/5}\;.
\end{equation}
We can derive how much faster this is with respect to $v_h$, the velocity corresponding to ballistic, homologous expansion:
\begin{equation}
    v_h = \frac{r}{t} \propto \frac{t^{6/5}}{t} = t^{1/5}\;; 
\end{equation} Which leads to 
\begin{equation}
\label{eq:v_before_spindown}
    v = \frac{6}{5} v_h,
\end{equation}
This equation assumes the power supplied by the wind to be constant. However, owing to spindown of the pulsar, the wind loses power. A significant fraction of the wind power is lost by the spindown age ($\tau$) of the pulsar (see Equation 3 of \citealt{YC15} for the exact definition). This implies that the ejecta in the PWN should relax to homologous expansion after this timescale. Using Equation \ref{eq:v_before_spindown}, the velocity at $t=\tau$ is

\begin{equation}
\label{eq:r0_tau_rel}
    v = \frac{6}{5} \frac{r_0}{\tau},
\end{equation}

assuming the size of the PWN to be equal to $r_0$ at $t=\tau$. Thereafter, the PWN expands ballistically, that is, $r(t>\tau) = r_0 + v(t-\tau)$. Eliminating $r_0$ using Equation \ref{eq:r0_tau_rel}, this can be written as an expression for the velocity of the ejecta after the pulsar has started spinning down:

\begin{equation}
    \label{eq:v_after_spindown}
    v = \left(1-\frac{\tau}{6t}\right)^{-1} \frac{r}{t} = \left(1-\frac{\tau}{6t}\right)^{-1} v_h
\end{equation}

Equations \ref{eq:v_before_spindown} and \ref{eq:v_after_spindown} give us an analytical prediction of the ratio between the observed velocity $v$ and homologous expansion velocity $v_h$ in a PWN, given the spindown age $\tau$ is known. The ratio $v/v_h$ is also the ratio of the real age to the inferred age of the PWN ($v=r/t'=(t/t')(r/t)=(t/t')v_h$, where $t'$ is the inferred age assuming unaccelerated expansion). It can be seen from Equation \ref{eq:v_after_spindown} that $v/v_h$ approaches unity as the age of the pulsar increases ($\tau \ll t$), or the ejecta assumes homologous expansion after the pulsar has had sufficient time to spin down. 

We can also calculate $v/v_h$ using a 3D numerical PWN model. This model was developed using \texttt{Sprout} \citep{Mandal2023}, a finite volume moving mesh hydrodynamics code with second order accuracy. \texttt{Sprout} solves the equations of ideal, non-relativistic hydrodynamics on a Cartesian mesh that can expand self-similarly with time. The model is based on the idea that the Crab resulted from a SN with atypically low energy \citep{YC15}, and that the pulsar wind is still inside the dense, inner part of the stellar ejecta \citep{Hester2008}. This implies there is freely expanding ejecta beyond the visible Crab, 
for which some evidence was found via blueshifted UV absorption lines, albeit accounting for only 0.3 M$_\odot$ of material.
(\citealt{Sollerman00}; but see also \citealt{Lund2012}). We start from a homologously expanding ejecta, with a mass of $5M_{\odot}$, and an ejecta energy of $10^{50}$ ergs. The ejecta density profile is chosen to be a double power law, with a constant density inner ejecta and steeply declining density ($\rho \propto r^{-9}$) in the outer ejecta. Calculations start at 1 month, and continue up to the present age of the Crab. The wind from the central pulsar is taken to be spherically symmetric, with the luminosity given by:

\begin{equation}
\label{eq:L_spindown}
    L(t) = L_0 \left( 1 + \frac{t}{\tau} \right)^{-\frac{n+1}{n-1}},
\end{equation}

where $L_0 = 10^{39}\mathrm{\;ergs/s}$, the pulsar braking index $n=2.51$ \citep{Lyne2015} and $\tau=690\mathrm{\,yrs}$ (using Equation 3 of \citealt{YC15} and parameters of the Crab pulsar from \citealt{Lyne2015}). These parameters ensure that the pulsar has spun down significantly at the present age of the model, and that the wind still remains confined by the constant density inner ejecta.  We obtain proper motion of the shocked plasma in the model in the plane of the sky as a function of the radial coordinate in the sky (similar to Figure 6). This data provides $v/v_h$ as a function of the pulsar age $t$. 

\begin{figure}
\centering
\includegraphics[width=3.3in]{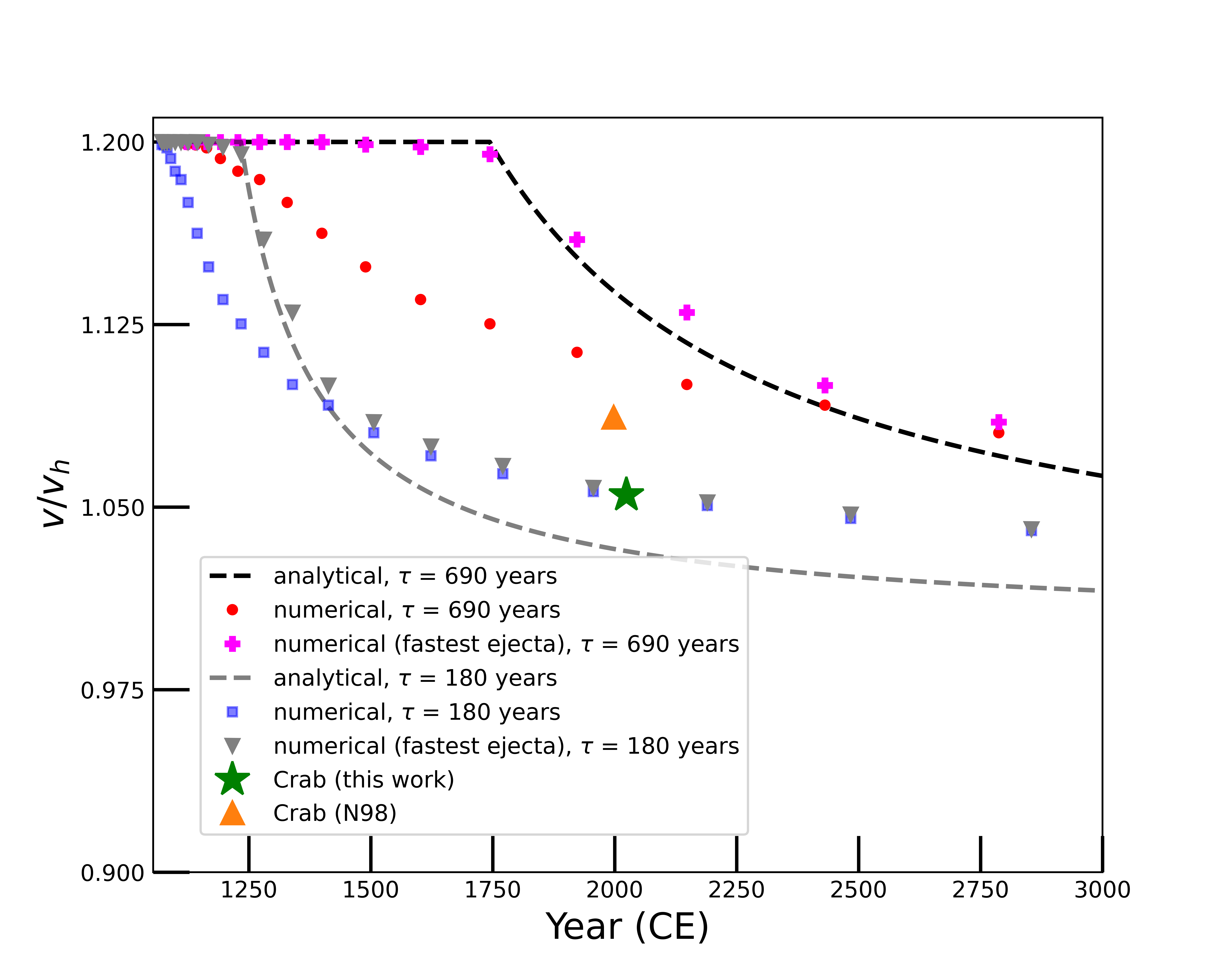}
\caption{The ratio of the observed velocity $v$ and ballistic homologous expansion velocity $v_h = r/t$ (also the ratio of the real age to the inferred age of the PWN) as a function of time. The green star shows the mass-unweighted average measured in this paper.}
\label{fig:k_vs_t}
\end{figure}

The analytical and numerical values of $v/v_h$ are shown in Figure \ref{fig:k_vs_t}, along with the average value of 
$v/v_h=1.06$ we derive for the Crab from our proper motion measurements. Note that we don't attempt to measure $v/v_h$ using the closest approach method because the position of the proper motion vectors cannot be traced back to the same origin and therefore Equations~\ref{eq:v_before_spindown} or~\ref{eq:v_after_spindown} do not remain valid.

Figure \ref{fig:k_vs_t} indicates the Crab nebula has relaxed closer to homologous expansion than indicated by the analytical or numerical models. 
Thus, the inner ejecta confined pulsar wind model of \citet{YC15}, with a spindown age of $690$ years, fails to explain the observed proper motion in the Crab. 

One way to interpret this discrepancy is that the wind has already blown through the inner ejecta and has made its way to the steeply dropping outer ejecta, as suggested by \citet{BC17}. In this phase, the pulsar wind nebula no longer remains self-similar, since hydrodynamic instabilities produce cracks or channels in the shocked shell, and the pulsar wind escapes through these channels. This could slow down the filaments more than in the self-similar case, since the wind blows through the shocked shell and cannot accelerate it as effectively as before. 

In such a case, one would expect the wind (emitting synchrotron radiation at radio wavelengths) to have a considerably larger $v/v_h$ ratio than the dense filaments seen in the optical. In fact, \citet{Bietenholz2015} measured multiple epochs of VLA 5 GHz radio images of the Crab and calculated  $v/v_h=1.26$ for the Crab synchrotron nebula (see Table~\ref{tab:centers}). If we consider the fast moving shocked plasma closest to the forward shock in our numerical model to be representative of the shocked wind and calculate $v/v_h$ only for this plasma, we obtain a curve that almost exactly matches the analytical prediction (Equations \ref{eq:v_before_spindown} and \ref{eq:v_after_spindown}) in Figure \ref{fig:k_vs_t}. Thus, at Crab's present age, we should expect a $v/v_h$ close to 1.15. The fact that it is greater than 1.2 is suggestive of a blowout, or perhaps a shallow density profile of the inner ejecta (say $\rho \propto r^{-1}$) instead of constant density.

We also explored an alternative model, where the spindown age is assumed to $180$ yr (blue squares in Figure \ref{fig:k_vs_t}), which is much less than the typically adopted age of $690$ yr, and higher initial luminosity. This is motivated in part by recent modelling of historical and contemporary lightcurves of the Crab that consider a significantly smaller spindown age of the pulsar \citep{Omand2024}. Although typical estimates of the spindown age are based on the assumption that the spin period of a pulsar evolves as a power-law (with a constant index) after it starts spinning down (see \citealt{YC15}), and this power-law index has been observed to be fairly constant ($\sim 2.51$) for the past 50 years \citep{Lyne1998,Lyne2015},  \citet{Lyne2015} show that the Crab exhibits `glitches' which are essentially large but short-duration deviations from a simple power law evolution.

We find close overlap between the $v/v_h$ of our $180$\,yr model and the proper motion-derived estimate reported in this paper. Nonetheless, many caveats are noted for this interpretation. Firstly, the proper motion measurements from our numerical model take into account all of the shocked plasma, yet it is unclear if all the shocked plasma in the Crab is bright enough to be included in the presented optical proper motion measurements. This is a factor that could influence the curve for the numerical model. Secondly, the wind from the Crab pulsar is not symmetric but equatorial and also includes a bipolar jet (see for example x-ray images of the Crab; \citealt{Weisskopf2000} and \citealt{Mizuno2023}). Likewise, the environment into which the supernova expanded into may have included a circumstellar disc \citep{Fesen1992,Smith13}. It is possible that the ratio $v/v_h$ will evolve somewhat differently for such a system. Furthermore, reproducing the radio convergence date reported in \citet{Bietenholz2015} would require adoption of a different ejecta profile (e.g., $\rho \propto r^{-1}$). Finally, the model assumes that the PWN is still expanding into the flatter part of the ejecta density profile, while the observed filament distribution is more consistent with the ``blowout'' scenario in which the PWN has reached the steep part of the ejecta profile \citep[e.g.,][]{Temim24}. This investigation is being pursued and will be published in a future study (Mandal et al., in prep). 

\section{Conclusions}
%%--------------------
 
We have presented proper motion measurements of the Crab Nebula made from three epochs of CFHT images spanning twelve years. Our data set of 19974 proper motion vectors with uncertainty $<10$\,mas\,yr$^{-1}$ spanning the majority of the Crab Nebula is the most extensive and detailed mapping of the Crab's two-dimensional expansion properties to date. The major findings of our analysis are as follows:

\begin{itemize}
    \item We estimate the centre of expansion of the Crab Nebula to be located at 5$^{h}$34$^{m}$32.67$^{s}$ +22$^{\circ}$00$^{\prime}$46.7$^{\prime\prime}$ (J2000), with 1$\sigma$ uncertainty of 1.2$^{\prime\prime}$. This origin is broadly consistent with previous estimates (e.g., \citealt{Trimble1968}). 
    \item We calculate the average date of convergence of the ejecta to be 1105.5 CE $\pm$ 0.5. Our date is 15-35 years earlier than most previous estimates, but still consistent with pulsar wind-driven acceleration of ejecta from a historically-recorded supernova explosion date of 1054. Our convergence date is consistent with the optical convergence date reported in \citet{Bietenholz2015}, but our estimate more faithfully traces the entire remnant using a uniform data set and suffers from less uncertainty.
    \item We show that the convergence date varies with position angle around the Crab Nebula, and that the greatest age associated with lowest acceleration is measured in the equatorial region aligned with the east-west synchrotron bays. This finding is consistent with an SN ejecta and PWN expansion into a disk-like CSM geometry.
    \item Comparing our results to previous analytical predictions of the expansion of the Crab Nebula suggests that the ejecta are closer to homologous expansion now than previously thought. We conducted a 3D numerical simulation of this expansion assuming a typically-adopted spindown age of 690 yr, and verified that the expected $v/v_h$ is higher than our observed value. Multiple lines of evidence, including the difference in convergence dates for the optical vs.\ radio-emitting components of the Crab \citep{Bietenholz2015}, favor that the discrepancy is a consequence of the pulsar wind having already blown through the inner flat part of the ejecta density profile \citep{BC17}. We also explore an alternative possibility that a more rapid spindown age of 180 yr may be more appropriate for the Crab pulsar, but note many caveats to this interpretation. 
\end{itemize} 

Accurate interpretation of the optical convergence date we estimate depends critically on the convergence date estimated from the radio-emitting pulsar wind radiation \citep{Bietenholz2015}. An independent measurement of the synchrotron nebula's proper motion and associated convergence date, made via optical or infrared imaging and closely following the asymmetry of the wind, would provide additional valuable information about the origin of the Crab Nebula's expansion dynamics. Also desirable are multiple epochs of SITELLE observations, which would permit proper motion investigations that can mitigate the confusion of multiple velocity components along the line of sight. Such a 3D expansion mapping would enable a more precise 3D rendering of the Crab Nebula than \citet{Martin2021}, and have the potential to robustly characterize the surrounding CSM geometry. 

\section*{Acknowledgements}

We are thankful to Marianne Ruest which pointed out the existence of
PIV algorithms, and to William Blair who read an earlier draft of the manuscript and provided helpful comments.

We are also thankful to the
\texttt{python} \citep{VanRossum2009} community and the free
softwares that made the analysis of this data possible: \texttt{numpy}
\citep{Oliphant2006}, \texttt{scipy} \citep{Virtanen2020},
\texttt{pandas} \citep{McKinney2010}, \texttt{panda3d}
\citep{Goslin2004}, \texttt{pyvista} \citep{Sullivan2019},
\texttt{matplotlib} \citep{Hunter2007} and \texttt{astropy}
\citep{Price2018}.

Based on observations obtained with MegaPrime/MegaCam, a joint project of CFHT and CEA/DAPNIA, at the Canada-France-Hawaii Telescope (CFHT) which is operated by the National Research Council (NRC) of Canada, the Institut National des Science de l'Univers of the Centre National de la Recherche Scientifique (CNRS) of France, and the University of Hawaii. The observations at the Canada-France-Hawaii Telescope were performed with care and respect from the summit of Maunakea which is a significant cultural and historic site. This paper is also based on observations obtained with SITELLE, a joint
project of Universit{\'e} Laval, ABB, Universit{\'e} de Montr{\'e}al
and the CFHT, the Institut National des
Science de l'Univers of the Centre National de la Recherche
Scientifique (CNRS) of France, and the University of Hawaii. 

D.M.\ acknowledges NSF support from grants PHY-2209451 and AST-2206532. LD is grateful to the Natural Sciences and Engineering Research Council of Canada and the Fonds de Recherche du Qu{\'e}bec for funding. T.T.\ acknowledges support from the NSF grant AST-2205314 and the
NASA ADAP award 80NSSC23K1130.

This work has made use of data from the European Space Agency (ESA)
mission {\it Gaia} (\url{https://www.cosmos.esa.int/gaia}), processed
by the {\it Gaia} Data Processing and Analysis Consortium (DPAC,
\url{https://www.cosmos.esa.int/web/gaia/dpac/consortium}). Funding
for the DPAC has been provided by national institutions, in particular
the institutions participating in the {\it Gaia} Multilateral
Agreement.

% The Acknowledgements section is not numbered. Here you can thank helpful
% colleagues, acknowledge funding agencies, telescopes and facilities used etc.

% %%%%%%%%%%%%%%%%%%%%%%%%%%%%%%%%%%%%%%%%%%%%%%%%%%

% %%%%%%%%%%%%%%%%%%%% REFERENCES %%%%%%%%%%%%%%%%%%

% % The best way to enter references is to use BibTeX:

\bibliographystyle{mnras}
\bibliography{m1} % if your bibtex file is called example.bib

% % Alternatively you could enter them by hand, like this:
% % This method is tedious and prone to error if you have lots of references
% \begin{thebibliography}{99}
% \bibitem[\protect\citeauthoryear{Author}{2012}]{Author2012}
% Author A.~N., 2013, Journal of Improbable Astronomy, 1, 1
% \bibitem[\protect\citeauthoryear{Others}{2013}]{Others2013}
% Others S., 2012, Journal of Interesting Stuff, 17, 198
% \end{thebibliography}

% %%%%%%%%%%%%%%%%%%%%%%%%%%%%%%%%%%%%%%%%%%%%%%%%%%

% %%%%%%%%%%%%%%%%% APPENDICES %%%%%%%%%%%%%%%%%%%%%

% If you want to present additional material which would interrupt the flow of the main paper,
% it can be placed in an Appendix which appears after the list of references.
\appendix

\section{Raw images registration}
\label{sec:megacam_images}

\begin{figure*}
  \includegraphics[width=\linewidth]{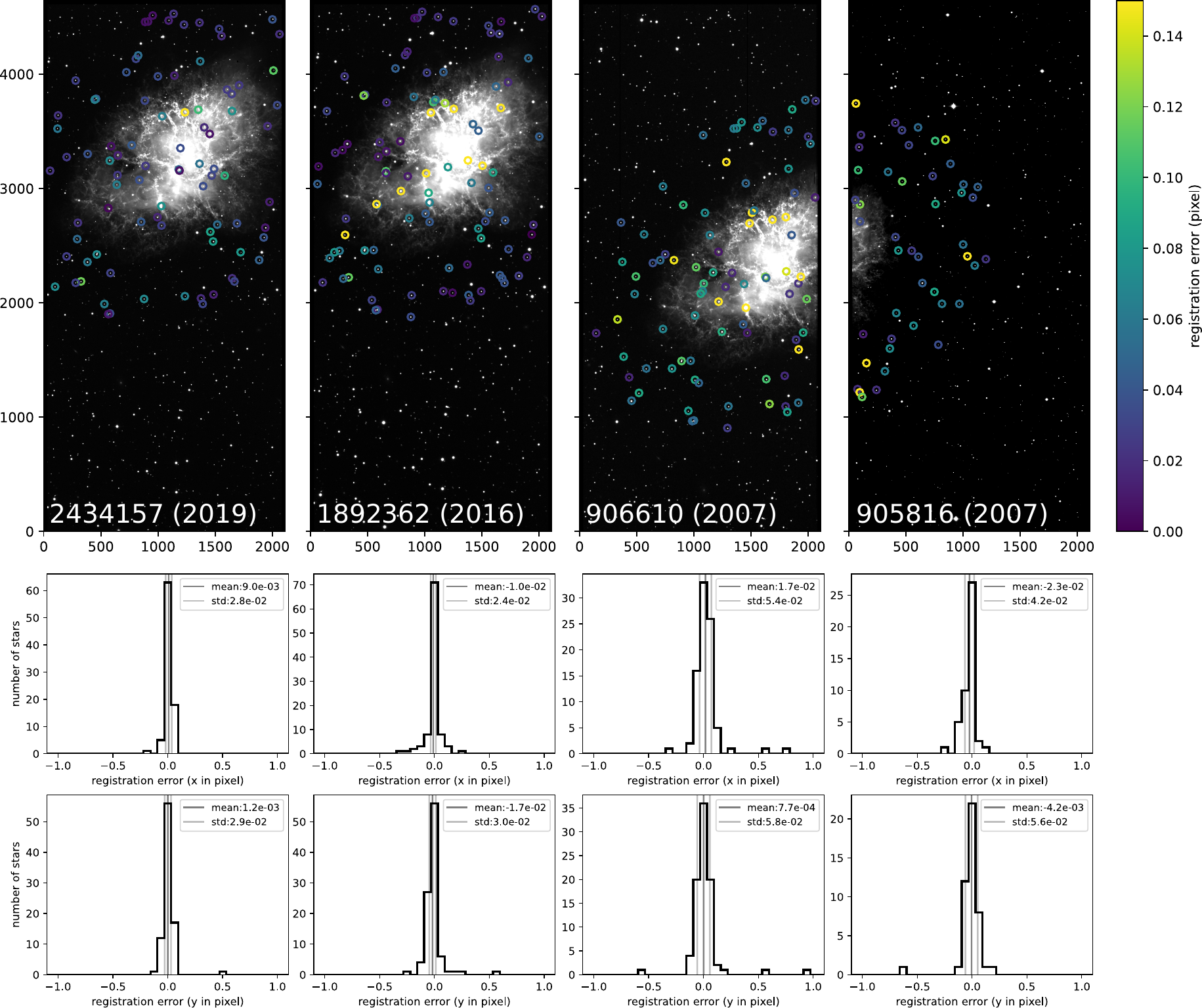}
  \caption{Images compared to compute the proper motion. The positions
    of the reference stars used for registration are shown as well as
    the registration error for each one of them. The distribution of
    the registration error along the x and y axes (roughly aligned
    with resp. the right ascension and declination) are shown in the
    panels below. The mean and standard deviation of the distribution
    are indicated.}
  \label{fig:chips}
\end{figure*}

%%%%%%%%%%%%%%%%%%%%%%%%%%%%%%%%%%%%%%%%%%%%%%%%%%

% Don't change these lines
\bsp	% typesetting comment
\label{lastpage}
\end{document}